\documentclass[useAMS,usenatbib]{mn2e}
\usepackage{natbib}
\usepackage{ graphicx}
\usepackage{ subfigure}
\usepackage{multirow}
\usepackage{amsmath}
\usepackage{amssymb}

\title[Black hole clustering]{Black hole clustering and duty cycles in the Illustris simulation}
\author[DeGraf \& Sijacki]  {C. DeGraf$^{1}$, D. Sijacki$^{1}$ \\
{1} {Institute of Astronomy and Kavli Institute for Cosmology, University of Cambridge, Madingley Road, Cambridge CB3 0HA, UK} 
}
\def\simgt{\lower.5ex\hbox{$\; \buildrel > \over \sim \;$}}

\begin{document}

\date{Submitted to MNRAS}
\pubyear{2016}

\maketitle
\begin{abstract}
We use the high-resolution cosmological simulation \textit{Illustris} to investigate the clustering of supermassive black holes across cosmic time, the link between black hole clustering and host halo masses, and the implications for black hole duty cycles.  Our predicted black hole correlation length and bias match the observational data very well across the full redshift range probed.  Black hole clustering is strongly luminosity-dependent on small, 1-halo scales, with some moderate dependence on larger scales of a few Mpc at intermediate redshifts. We find black hole clustering to evolve only weakly with redshift, initially following the behaviour of their hosts. However below $z \sim 2$ black hole clustering increases faster than that of their hosts, which leads to a significant overestimate of the clustering-predicted host halo mass.  The full distribution of host halo masses is very wide, including a low-mass tail extending up to an order of magnitude  below the naive prediction for minimum host mass.  
Our black hole duty cycles, $f_{\rm{duty}}$, follow a power-law dependence on black hole mass and decrease with redshift, and we provide accurate analytic fits to these.  The increase in clustering amplitude at late times, however, means that duty cycle estimates based on black hole clustering can overestimate $f_{\rm{duty}}$ substantially, by more than two orders of magnitude.  We find the best agreement when the minimum host mass is assumed to be $10^{11.2} M_\odot$, which provides an accurate measure across all redshifts and luminosity ranges probed by our simulation.

\end{abstract}

\section{Introduction}
\label{sec:intro}

It is now widely understood that supermassive black holes are found at the centre of massive galaxies \citep{KormendyRichstone1995}, and that properties of the host galaxy strongly correlate with black hole mass \citep[e.g.][]{Magorrian1998, Gebhardt2000, Graham2001,Ferrarese2002, Tremaine2002, HaringRix2004, Gultekin2009, McConnellMa2013, KormendyHo2013}. 

One fundamental aspect of black hole studies is clustering behaviour, which provides a unique way of linking black holes to their host galaxies.  Black hole clustering has been studied extensively in observations \citep[e.g.][]{LaFranca1998, Porciani2004, Croom2005, Shen2007, Myers2007, daAngela2008, Shen2009, Ross2009, White2012, Ikeda2015, Eftekharzadeh2015}, as well as simulations \citep[e.g.][]{Bonoli2009, Croton2009, DeGraf2010, DeGraf2012}.

Since the emergence of large scale surveys capable of probing a range of redshifts, the general consensus is that the clustering signal decreases with time.  At low redshift (below $z \sim 2$) the evidence for redshift evolution is generally weak, but at higher redshifts the evidence is much stronger, with correlation lengths approaching $10$ h$^{-1}$ Mpc \citep[e.g.][]{Myers2006, White2012}
and bias factors (the clustering strength relative to that of the underlying dark matter density distribution) as large as $b = 5-10$ \citep[e.g.][]{Shen2009, Ikeda2015}.
In addition to the redshift evolution, the possibility of luminosity dependent clustering has crucial implications for our understanding of the relation between black holes and their host halos.  In particular, under the simplistic assumption that Active Galactic Nuclei (AGN) luminosity is proportional to host halo mass, one would expect brighter samples to be more strongly clustered (consistent with the stronger clustering of more massive halos).  On the other hand, most models suggest a more widely varying black hole luminosity history, such that both bright and faint AGN can populate similar halos at different phases of their lifetimes, in which case clustering behaviour should only weakly depend on instantaneous luminosity.  
Many observations have found a lack of luminosity dependence \citep[e.g.][]{Croom2005, Myers2007, daAngela2008, White2012, Krolewski2015} or only a weak dependence \citep[e.g.][]{Shen2009, Eftekharzadeh2015}, supporting this model.  Work by \citet{Bonoli2009} suggests, however, that even in the case of varying luminosity histories, a luminosity dependence could be found among lower-luminosity black holes, which simulations are well suited to investigate as observations being to push to lower flux limits.

By matching quasar clustering to that of dark matter halos, the typical mass of the halos that host quasars can be estimated
, providing a relatively simple means of estimating host properties for a range of black hole populations.  
By taking the expected number density of such halos and combining with the number density of AGN (via a luminosity function), one can estimate the active fraction of black holes or duty cycle \citep[see, e.g.][]{HaimanHui2001, MartiniWeinberg2001, Grazian2004, Shankar2010}.
Such duty cycle estimates, however, rely upon several assumptions, most significantly the accuracy of the typical and/or minimum host halo mass calculated from clustering.  Furthermore, these estimates rely upon the link between AGN luminosity and the host halo.  Some models, however, suggest that only peak AGN luminosity correlates with host halo mass \citep[e.g.][]{Hopkins2005,Hopkins2005c}; in these models the scatter between low-luminosity lifetimes and host properties suggests that clustering should have a weaker luminosity dependence \citep[e.g.][]{Lidz2006} and that the assumptions used when estimating duty cycles may not hold.  Large-scale cosmological simulations are well suited for investigating these aspects of black hole clustering, which we focus on in this paper.

Here we use the state-of-the-art Illustris simulation \citep{Nelson2015} to study the clustering of supermassive black holes across cosmic time, taking advantage of the statistically representative sample provided by a large-volume simulation.  The Illustris simulation is a (106.5 Mpc)$^3$ box, providing a sufficiently large sample to predict clustering behaviour in detail for $z = 0-4$, including dependence on black hole luminosity.  The Illustris simulation has been shown to reproduce several key black hole properties, including black hole mass density, mass function, luminosity function, and scaling relations \citep{Sijacki2015}, making it ideally suited to investigate black hole clustering behaviour.  In addition to showing the clustering via the black hole autocorrelation function, we compare the clustering amplitude via both the correlation length and bias parameters to observational results. Using the strength of the black hole clustering signal, we are able to estimate the typical mass of host halos similar to observational approaches, and directly compare to the actual distribution of host masses and explain the discrepancies therein.  Similarly, clustering can be used to estimate the duty cycle of black holes \citep[see, e.g.][]{Eftekharzadeh2015}, which we compare directly to the actual duty cycle in the simulation, probing the accuracy of this estimate for several different definitions for duty cycle. 

The outline for our paper is as follows.  In Section \ref{sec:method} we outline the numerical methods used, including the Illustris simulation project, and the clustering calculation used throughout the paper.  In Section \ref{sec:results} we discuss the results of our investigation.  Section \ref{sec:bhclustering} covers the clustering of black holes, their evolution with redshift, and dependence on black hole luminosity.  In Section \ref{sec:hosthalo} we link the clustering behaviour to properties of host halos. In Section \ref{sec:dutycycle} we characterize the duty cycle of both black holes and halos, and the issues involved in estimated duty cycle from clustering behaviour.  Finally, we summarize our conclusions in Section \ref{sec:conclusions}.

\section{Method}
\label{sec:method}
\subsection{Simulations}
\label{sec:numerical}

In this study we analyse the Illustris\footnote{http://www.illustris-project.org; \citet{Nelson2015}} suite of simulations performed using the hydrodynamical code AREPO \citep{Springel2010}.  This code uses a TreePM gravity solver and a second-order unsplit Godunov method for solving for the hydro forces.  The hydrodynamics equations are solved on an unstructured Voronoi mesh which can move with the fluid in a quasi-Lagrangian manner.  Numerous computational and cosmological tests have been performed on the code, verifying its ability to properly capture shock properties, develop fluid instabilities and maintain low numerical diffusivity and Galilean invariance \citep[see, e.g.,][]{Springel2010, Springel2011, BauerSpringel2012, Sijacki2012, Vogelsberger2012, Keres2012, Torrey2012, Nelson2013}.

The Illustris suite of simulations \citep{Vogelsberger2014a, Genel2014} analyzed here uses a $(106.5 \rm{Mpc})^3$ cosmological box at several resolutions, with dark-matter only, non-radiative, and full hydrodynamic runs.  For this work, we focus only on the full-hydro high-resolution simulation (Illustris-1).  This run uses a standard $\Lambda$CDM cosmology, with $\Omega_{m,0}=0.2726$, $\Omega_{\Lambda,0}=0.7274$, $\Omega_{b,0}=0.0456$, $\sigma_8=0.809$, $n_s=0.963$, $H_0=70.4 \, \rm{km}\, \rm{s}^{-1} \, \rm{Mpc}^{-1}$ \citep[consistent with][]{Hinshaw2013} and runs from $z_{\rm{start}}=127$ to $z_{\rm{end}}=0$.  The simulation has $3 \times 1820^3$ resolution elements with typical gas cell mass $m_{gas}=1.26 \times 10^6 M_\odot$ and gravitational softening $\epsilon_{\rm{gas}}=0.71$ kpc and dark matter particle mass $m_{DM}=6.26 \times 10^6 M_\odot$ and gravitational softening $\epsilon_{\rm{DM}}=1.42$ kpc.

The Illustris simulations include a detailed model of the physics involved in galaxy formation.  
Primordial and metal-line cooling are included in the presence of a time-dependent UV background \citep{FaucherGiguere2009} including self-shielding \citep{Rahmati2013}; star formation and associated supernovae feedback follow the model of \citet{SpringelHernquist2003}, using a softer equation of state \citep{Springel2005} with $q = 0.3$ and a Chabrier initial mass function \citep{Chabrier2003}.  Models for stellar evolution, gas recycling and metal enrichment are also included \citep[see also][]{Wiersma2009}, along with mass- and metal-loaded galactic outflows \citep[see also][]{OppenheimerDave2008, Okamoto2010, PuchweinSpringel2013}.  

Black holes are treated as collsionless sink particles.  Seeding is based on an on-the-fly Friends-of-Friends algorithm, where black hole particles with seed mass of $10^5 \: h^{-1} \: M_\odot$ are inserted into halos with mass above $5 \times 10^{10} \: h^{-1} \: M_\odot$ which do not already contain a black hole particle.  This seeding prescription is intended to remain consistent with a variety of formation models, including direct collapse of gas clouds \citep[e.g.][]{BrommLoeb2003, Begelman2006}, or formation of smaller seeds from early PopIII stars \citep{BrommLarson2004, Yoshida2006} followed by a period of rapid growth leading to our seed mass. After seeding, black holes grow though accretion via gas accretion and black hole mergers, which occur when a pair of black holes pass within their respective smoothing lengths.  Gas accretion is based on a Bondi-Hoyle-like formalism \citep[$\dot{M}_{\rm{BH}} = (4 \pi \alpha G^2 M_{\rm{BH}}^2 \rho)/c_s^3$][]{BondiHoyle1944, Bondi1952}, with an imposed upper limit of the Eddington rate [$\dot{M}_{\rm{Edd}}=(4 \pi G M_{\rm{BH}} m_p) / (\epsilon_r \sigma_T c)$].  In addition, a pressure criterion is applied to lower the accretion rate if the ambient gas pressure is insufficient to compress the gas to above the star formation density threshold, preventing formation of unrealistic bubbles in low-density gas \citep[see][ for more details]{Vogelsberger2013}.

Black hole feedback is included in three separate modes: ``quasar'', ``radio'' and ``radiative.''  In quasar mode, feedback is radiated with an efficiency of $\epsilon_r = 0.2$, and couples thermally to the surrounding gas with an efficiency $\epsilon_f = 0.05$.  For black holes with accretion efficiency below $\chi_{\rm{radio}} = \dot{M}_{\rm{BH}}/\dot{M}_{\rm{Edd}} = 0.05$, the radio mode is used.  Radio mode feedback is applied by inserting energy into hot bubbles randomly placed around the black hole, representing bubbles expected to be inflated by AGN radio jets, with the bubble energy set by $E_{\rm{bubble}} = \epsilon_m \epsilon_r \Delta M_{\rm{BH}} c^2$, with a coupling factor $\epsilon_m = 0.35$.  Finally, radiative feedback is incorporated by modifying the photo-ionization and photo-heating rates near accreting black holes.  For a more detailed discussion of the black hole accretion and feedback model, see \citet{Sijacki2007, Sijacki2015}.  

We do note several uncertainties regarding black hole modeling as implemented here.  First, the seeding mechanism used is based solely on host halo mass, and does not attempt to characterize the physical process behind formation.  Different formation pathways can lead to dramatically different initial mass scales (and thus early accretion rates and associated luminosities), ranging from low-mass seeds from PopIII stars to more massive seeds from runaway interactions in dense nuclear star clusters or direct collapse of massive gas clouds.  Although the model used here is intended to be consistent with any of these, the different environmental factors associated with each formation mechanism could potentially be imprinted on black hole clustering.  However, this effect would be strongest among very low-mass black holes and would likely not have a significant impact on the scales considered in this paper (though future work will investigate dependencies on black hole seed formation).  We also note that the accretion and feedback models used here are limited by the resolution possible in a large volume simulation.  In particular, we note that different driving forces behind black hole fueling could potentially be found in the clustering signal; e.g. the different environments in which one might find merger-induced fueling \citep[e.g.][]{DiMatteo2005} versus instability driven growth \citep[e.g.][]{GaborBournaud2014}.  However, computational limitations currently prevent resolving the formation of such instabilities in sufficient volume to study clustering \citep[though see][]{DeGraf2016clumpy}.  Investigations into these areas may provide the means by which to better distinguish mechanisms for formation and growth of supermassive black holes, but are beyond the scope of this paper.  For this investigation we focus on the more general model for black holes, which has been demonstrated to accurately reproduce a wide range of black hole properties and the correlation with host galaxies \cite[see][]{Sijacki2015}.

For further details on the Illustris simulations, see \citet{Vogelsberger2014a,Vogelsberger2014b, Genel2014, Sijacki2015}.

\begin{figure*}
    \centering
    \includegraphics[width=16.0cm]{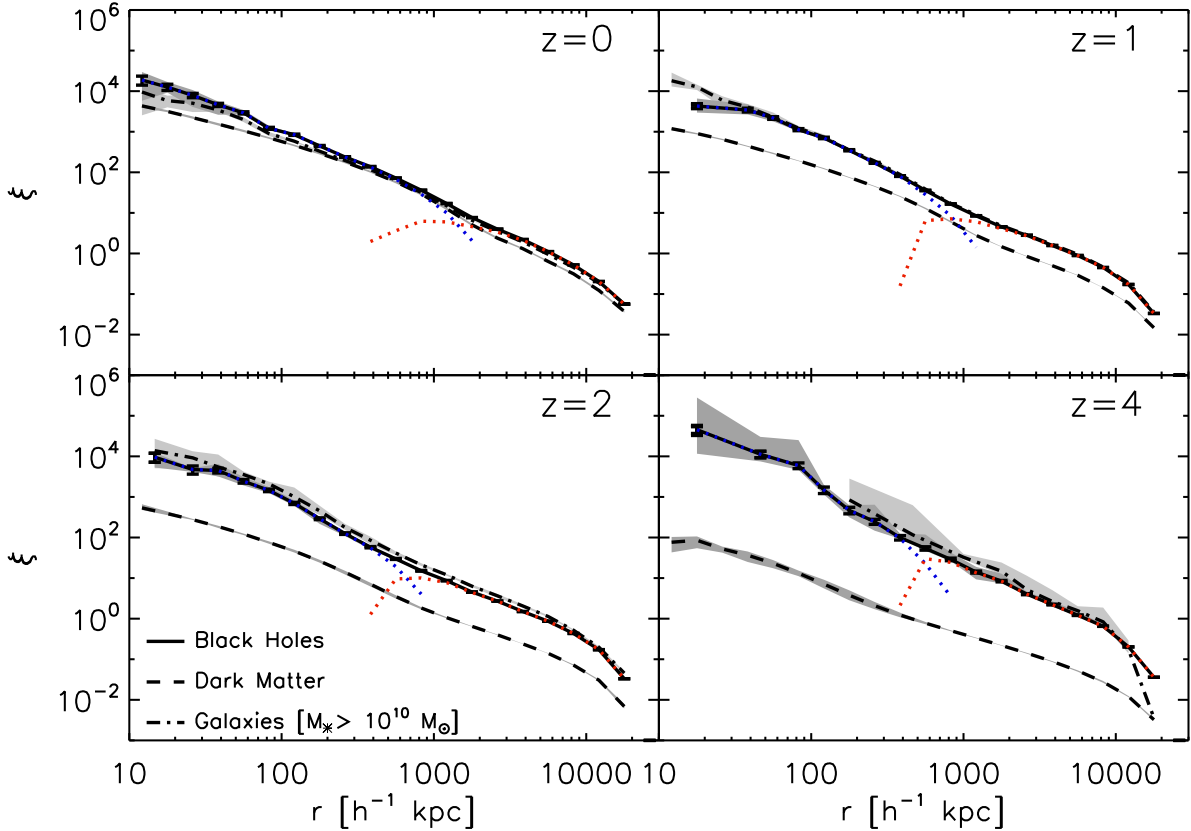}
    \caption{Autocorrelation for black holes (solid line; no luminosity cut and with Poisson error bars), dark matter (dashed line), and galaxies ($M_* > 10^{10} M_\odot$; dot-dashed line) at $z$ = 0, 1, 2 and 4.  The shaded region represents the variation across five sequential snapshots (at $\sim$150 Myr per snapshot), characterizing the uncertainty in the relation.  The blue and red dotted lines are the 1-halo and 2-halo components of the black hole correlation function.   The black hole correlation function evolves significantly slower than the dark matter, such that the bias between them decreases with redshift.}
    \label{fig:autocorrelation}
\end{figure*}

\subsection{Clustering calculations}
\label{sec:clustering}

To characterize the clustering of black holes in our simulation, we use the black hole autocorrelation function $\xi$ and the correlation length $r_0$ (the scale at which $\xi (r_0) = 1$).  The correlation function is calculated using the natural estimator
\begin{equation}
\begin{split}
\xi (r) &= \frac{DD}{RR}-1 \\
&= \frac{DD}{N_{\rm{obj}}(N_{\rm{obj}}-1) \frac{\Delta V}{V}}-1
\end{split}
\label{eq:basexi}
\end{equation}
where DD is the number of data pairs with a separation of $r \pm \Delta r/2$, RR is the number of pairs expected from a random distribution, $N_{\rm{obj}}$ is the number of objects in the data set, $\Delta V$ is the volume of a spherical shell with radius $r$ and thickness $\Delta r$, and $V$ is the volume of the simulation box.  We note that the periodicity of our simulation boundaries means that edge effects are minimal, confirmed by comparisons with the \citet{LandySzalay1993} estimator which provides consistent results. 
For calculation of $\xi_{\rm{DM}}$ (plotted in Figure \ref{fig:autocorrelation} and used in Equation \ref{eq:bias}) we use a sample of $\sim$400,000 DM particles selected at random from the full simulation box, which provides results converged to within a few percent.

\section{Results}
\label{sec:results}
\subsection{Black hole clustering}
\label{sec:bhclustering}

In Figure \ref{fig:autocorrelation} we show the autocorrelation function of black holes (solid line), dark matter (dashed line), and galaxies ($M_* > 10^{10} M_\odot$; dot-dashed line) for $z$= 0, 1, 2 and 4, with shaded regions representing the variation across 5 sequential snapshots (at $\sim$150 Myr per snapshot).  For all but the smallest scales, the inter-snapshot variation is minimal, as expected.  We also include Poisson error bars for the black hole correlation function, which shows the Poisson errors are even smaller than the inter-snapshot variation.  The dark matter clustering evolves such that $\xi_{\rm{DM}}$ increases with time, as predicted by linear growth models (at least at large scales; at smaller scales non-linear growth dominates). 
The black hole clustering shows much weaker evolution with redshift, such that the bias between $\xi_{\rm{BH}}$ and $\xi_{\rm{DM}}$ decreases as we approach lower redshift, which we address more explicitly below.

In addition to the full correlation, we separate $\xi_{\rm{BH}}$ into 1-halo (dotted blue) and 2-halo (dotted red) components, defined similarly to Equation \ref{eq:basexi}: $\xi_{1h} = DD_{1h}/RR-1$ and $\xi_{2h} = DD_{2h}/RR-1$, where $DD_{1h}$ ($DD_{2h}$) are pairs of black holes found in the same (different) halos.  The crossover between 1-halo and 2-halo terms occurs at $\sim 1$ Mpc at $z$ = 0 and at progressively smaller scales for higher redshift, consistent with the expectation that the typical halos hosting black holes are larger at lower redshift.  We also note that the 2-halo term completely dominates at larger scales; for this reason we use scales above $\sim 2$ Mpc when calculating the best-fitting functions for correlation length and bias factors (see Section \ref{sec:hosthalo}).

\begin{figure*}
    \centering
    \includegraphics[width=15.0cm]{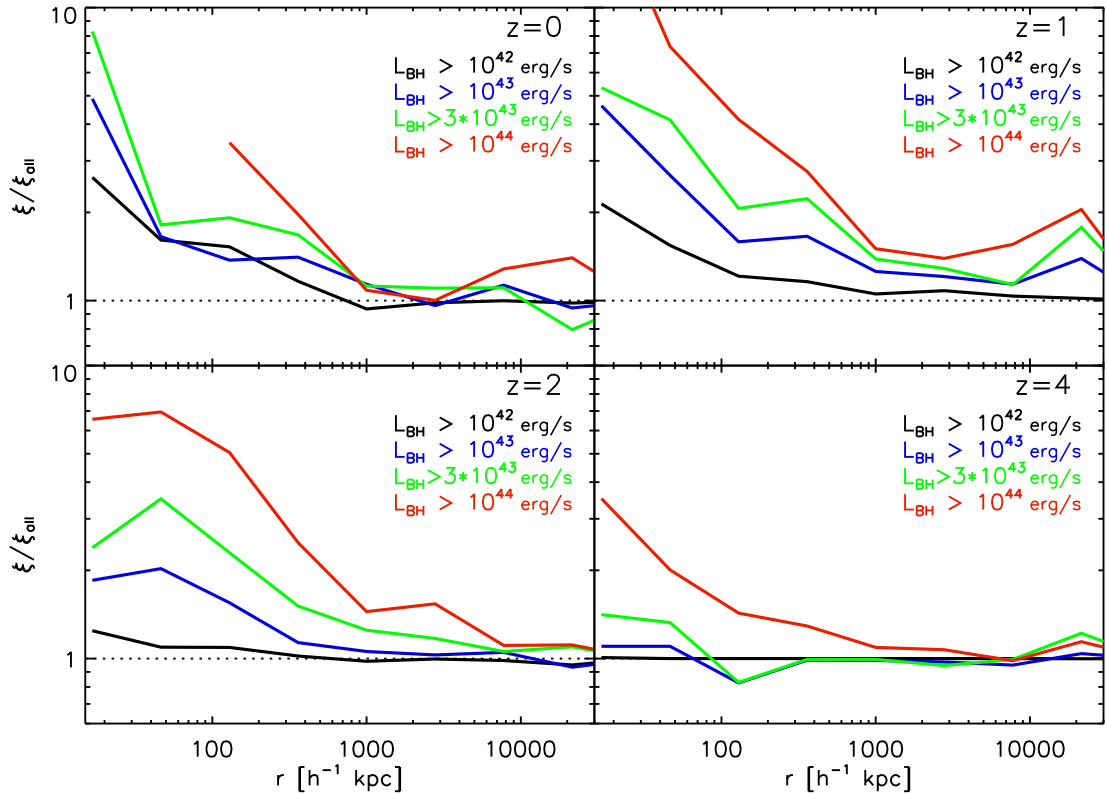}
    \caption{Ratio between autocorrelation functions using luminosity-selected population $\xi$ and full population $\xi_{\rm{all}}$.  For each curve, we take the median ratio between three sequential snapshots ($\sim$150 Myr per snapshot) to smooth out short-term variations.  We find the luminosity-dependent clustering is primarily at small, 1-halo scales, but does extend out to larger scales, especially at intermediate redshift.}
    \label{fig:lumdep}
\end{figure*}

To investigate the luminosity dependence of clustering behaviour, in Figure \ref{fig:lumdep} we show how progressively higher cuts on $L_{\rm{BH}}=\epsilon_r \dot{M} c^2$ affect $\xi_{\rm{BH}}$.  To more clearly show the effect, we plot the ratio between the correlation function of luminosity-selected samples and that of the full black hole population ($\xi ({\rm{L_{\rm{BH}}>L_{\rm{cut}}}})/\xi_{\rm{BH}}$).  We note that the luminosity dependence is strongest at intermediate redshift ($z \sim 1-2$), when AGN activity is quite high and before self-regulation has slowed black hole growth.  We also see that luminosity dependence is strongest at the smallest scales (well within the 1-halo term).  This suggests that more luminous black holes tend to be strongly clustered within individual halos (and thus strengthening the 1-halo term), which may be explained by more luminous AGN tending to be found in more massive halos, which in turn tend to host the largest number of satellite black holes necessary to produce a small-scale 1-halo signal \citep[see][]{Chatterjee2011,DeGrafHOD2011}.

\begin{figure}
    \centering
    \includegraphics[width=8.5cm]{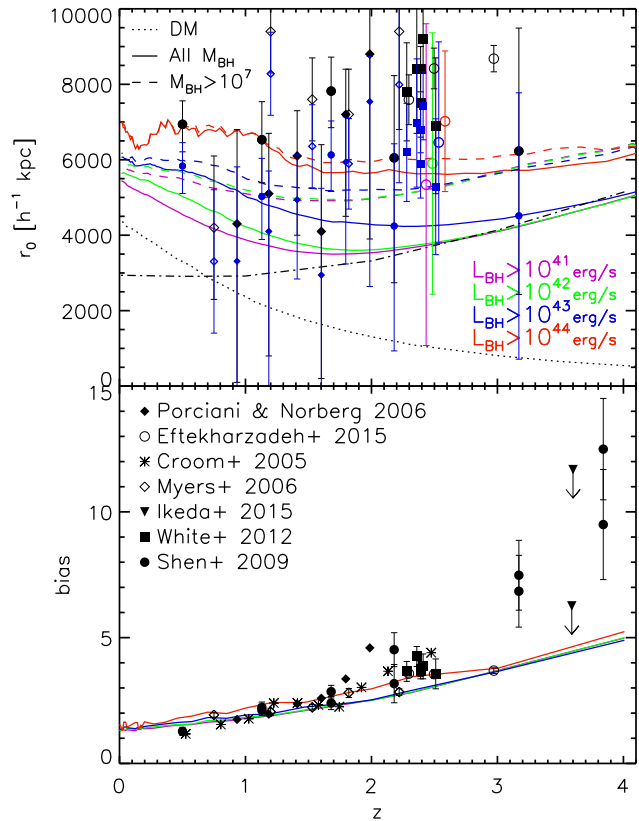}
    \caption{\textit{Top:} Correlation length of black hole autocorrelation function for several luminosity cuts.  Curves are smoothed to show overall trend rather than short-term variations in the high-luminosity curve.  \textit{Solid lines}: All black holes included.  \textit{Dashed lines}: Only black holes with $M_{\rm{BH}} > 10^7 M_\odot$ included.  \textit{Dotted lines:} Dark matter correlation function.  \textit{Dot-dashed line:} Halo correlation function for halos with $M_{\rm{DM}} > 10^{11.2} M_\odot$.  The blue coloured points represent observational data which has been adjusted according to the luminosity dependence of \citet{Eftekharzadeh2015} to match the mean luminosity of our $L_{\rm{BH}} > 10^{43}$ erg s$^{-1}$ sample.  We find the correlation length to be moderately-dependent on $L_{\rm{BH}}$, with generally weak evolution with redshift.  \textit{Bottom:}  Bias (relative to DM autocorrelation) for luminosity selected black holes, compared to observations.  Observational data is from \citet{Croom2005, PorcianiNorberg2006, Myers2006, Shen2009, White2012, Ikeda2015, Eftekharzadeh2015}, with bias values adjusted to account for differences in $\sigma_8$.  }
    \label{fig:corrlength}
\end{figure}

To better characterize the evolution of the large-scale black hole clustering, we use the correlation length ($r_0$), defined as the scale at which $\xi (r_0)$ = 1.  In Figure \ref{fig:corrlength} we show the correlation length for four luminosity-selected black holes samples.  We calculate $r_0$ using a power-law fit to $\xi$ in the range 2-10 Mpc (where the 2-halo term dominates), but note that the value is not sensitive to the exact fitting range selected.  Solid lines show black holes selected regardless of mass, while dashed lines only include black holes with $M_{\rm{BH}} > 10^7 M_\odot$.  For the most luminous black holes in our sample ($L_{\rm{BH}} > 10^{44}$ erg s$^{-1}$) we find roughly constant $r_0$, with only a slight increase at $z \sim 1-1.5$ (and minimal change when imposing the cut on $M_{\rm{BH}}$, since most luminous AGN are above $10^7 M_\odot$).  For fainter luminosity cuts we find the correlation length tends to decrease with time to a minimum at $z \sim 2$, followed by an increase at later times, and fainter black holes tend to be less-strongly clustered.  Similar to the semi-analytic work by \citet{Bonoli2009}, we find a significant luminosity dependence when considering a sufficiently large range of luminosities.  We also show the correlation length of dark matter halos with $M_{\rm{DM}} > 10^{11.2} M_\odot$, which closely matches the low-luminosity black hole clustering for $z > 2$, suggesting the occupation distribution remains roughly constant at high-redshift, consistent with earlier findings \citep{DeGraf2012}.  At lower redshifts, the increasing black hole correlation length suggests an increase in typical host halo mass, which we investigate in more detail in Section \ref{sec:hosthalo}.

We also compare with observational measurements of $r_0$, and find generally consistent measurements.  We note that our $r_0$ values are slightly lower than the observations, however this is at least in part due to the difference between the flux-limited surveys and our volume limited sample.  As an example, we show the observations from \citet{Eftekharzadeh2015} adjusted for luminosity (using their luminosity fit) as coloured circles.  Although our predictions are still lower than theirs, they are within the error bars, and closely match when considering only black holes with $M_{\rm{BH}} > 10^7 M_\odot$.  We also provide green datapoints, representing observational data adjusted to match the luminosity of our $L_{\rm{BH}} > 10^{43}$ erg s$^{-1}$ sample.  This adjustment compares the mean luminosity of the observed samples\footnote{Where necessary, we apply bolometric corrections based on the quasar SED of \citet{Hopkins2007}.} and our z- and $L_{\rm{BH}}$-selected samples, and applies the luminosity dependence formula from \citet{Eftekharzadeh2015}, producing good agreement between our predictions and observed data.  We do note that the $L_{\rm{BH}}$-dependence of \citet{Eftekharzadeh2015} is highly uncertain, however the $L_{\rm{BH}}$-evolution is consistent with our simulation (see coloured open circles in Figure \ref{fig:corrlength}).

Having shown that the Illustris simulations produce expected behaviour for the black hole autocorrelation functions, we investigate how the clustering behaviour links black hole properties to those of their host halos.  In the bottom panel of Figure \ref{fig:corrlength} we plot the evolution of the black hole bias, defined as 
\begin{equation}
b = \sqrt{\xi_{\rm{BH}}/\xi_{\rm{DM}}} \: .
\label{eq:bias}
\end{equation}
We calculate the bias over the range 2-10 Mpc, which keeps us above the 1-halo regime \citep[where non-linear bias can be significant; see Figure \ref{fig:autocorrelation} and][]{DeGrafClustering2010}, and below the scale at which box-size will be a limiting factor.  Observations from a range of studies have been included (see caption), having been adjusted to account for assumed $\sigma_8$ value. At low redshift (below $z \sim 2$) we have excellent agreement with observations.  At higher redshift, we appear to underestimate the bias.  However, we note that the disagreement is primarily from the \citet{Shen2009} results \citep[as the][ results are upper-limits only]{Ikeda2015}.  However, \citet{Eftekharzadeh2015} note that the \citet{Shen2009} measurement is dominated by a single bin at $\sim 35 \: h^{-1}$ Mpc, a scale well above that probed by our simulation.  
We find that the evolution of the black hole bias parameter as a function of redshift is well fit by a second-order polynomial $b(z)=A+Bz+Cz^2$.  We provide the best-fitting parameter values in Table \ref{tab:biasevolution}.

\subsection{Host halo properties}
\label{sec:hosthalo}

\begin{figure}
    \centering
    \includegraphics[width=8.5cm]{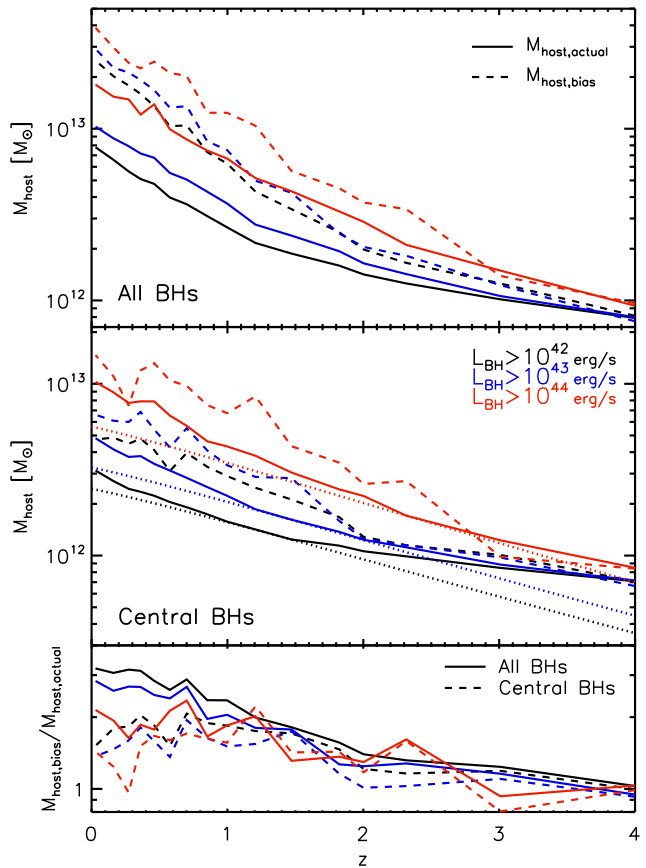}
    \caption{\textit{Top:} Mass of halos hosting black holes.  Solid lines show the actual mean mass (in log-space); Dashed lines show the predicted host mass based on the calculated bias. \textit{Middle:} Same as top, but using only central black holes. Dotted lines represent expected halo mass for the median halo growth fitting function of \citet{Fakhouri2010}, matching at the redshift where our mean halo mass begins increasing faster than the expected growth curve. \textit{Bottom:} Ratio between bias-predicted host mass and actual most mass, for all black holes (solid lines) and for central black holes only (dashed lines).  Black hole clustering increases faster than a $M_{\rm{host}}$-selected halo sample, leading to an overestimate in the clustering-predicted host halo mass.}
    \label{fig:hostevolution}
\end{figure}

\begin{table}
\centering
\begin{tabular}{|c|c|c|c|}
\hline
$L_{\rm{BH,min}}$ & A & B & C \\
\hline
$10^{42}$ & 1.343 & 0.315 & 0.153 \\
$10^{43}$ & 1.390 & 0.365 & 0.131 \\
$10^{44}$ & 1.468 & 0.605 & 0.083 \\
\hline
\end{tabular}
\caption{Best fit parameters for evolution of bias $b(z) = A + Bz + Cz^3$.}
\label{tab:biasevolution}
\end{table}

In the top panel of Figure \ref{fig:hostevolution} we estimate the typical mass of halos hosting the luminosity-selected black hole samples.  The solid lines show the actual host masses $\langle \rm{log} (M_{\rm{host}}) \rangle$, where we find the expected trend of increasing host mass with time.  The dashed lines show the predicted halo mass based solely on the black hole clustering, by matching black hole clustering to that of dark matter halos.  
To make this prediction, we calculate the clustering bias over 2-10 Mpc (as in Section \ref{sec:bhclustering}).  We use the 2-10 Mpc range to remain in the 2-halo dominated regime (as seen in Figure \ref{fig:autocorrelation}).  Although limited to 2-halo scales, we note that satellite black holes may nonetheless contribute to the bias calculation, which we discuss below.  We compare this to the halo bias from the formalism of \citet[][see Equation 6]{Tinker2010} adopting a linear matter variance ($\sigma(M)$) calculated using the power spectrum from CAMB\footnote{http://camb.info} (with the cosmological parameters used in the Illustris simulations).\footnote{We have compared the halo correlation function from Illustris to the prediction for equal-mass halos using this approach, and confirmed excellent agreement.}
We find typical host halos of $\sim 10^{12}-10^{13} M_\odot$, consistent with general observational estimates of a few times $10^{12} \: h^{-1} \: M_\odot$ \citep[e.g.][]{Myers2007, daAngela2008, Ross2009, White2012, Ikeda2015}.

\begin{figure}
    \centering
    \includegraphics[width=8.5cm]{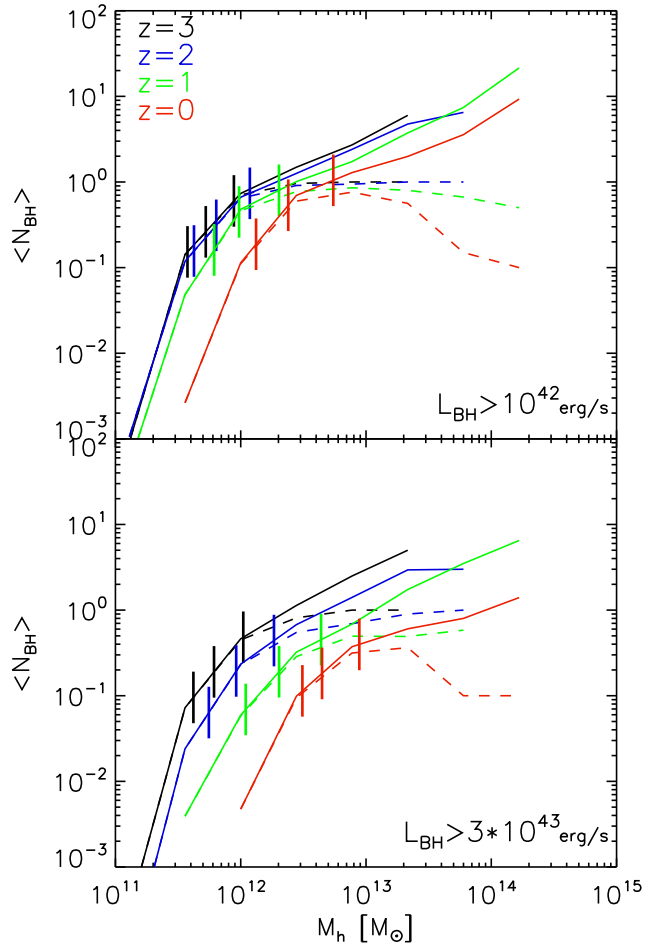}
    \caption{Mean occupation number of black holes above $10^{42}$ erg s$^{-1}$ (top) and $3 \times 10^{43}$ erg s$^{-1}$ (bottom).  Solid lines show full black hole population; dashed lines show central black holes only.  Vertical lines mark the host mass scale above which 90\%, 75\%, and 50\% of black holes are found.  Despite the lower occupation number, most black holes are found in the more common low-mass halos, and high-mass halos $\left (M_h \gtrsim 10^{12.5}-10^{13} M_\odot \right )$ tend to have significant numbers of satellite black holes.}
    \label{fig:occupationnumber}
\end{figure}

At high redshift (above $z \sim 2$) the agreement between the two actual mean host mass and the bias-predicted mass is good; for $z < 2$, however, the bias-prediction overestimates the typical host mass by a factor of $\sim$ 2.  At least part of this is due to halos hosting multiple black holes: larger halos tend to host larger numbers of satellite black holes, which biases $\xi$ toward the clustering of the larger (and thus more strongly clustered) halos.  We show this explicitly in Figure \ref{fig:occupationnumber}, where we plot the mean number of black holes per halo as a function of halo mass, showing that massive halos tend to host multiple black holes above a given $L_{\rm{BH,min}}$.  For comparison, the dashed lines in Figure \ref{fig:occupationnumber} show the mean number of \textit{central} black holes (defined as the most massive black hole in a given Friends-of-Friends defined halo) above $L_{\rm{BH,min}}$, which avoids this issue (note that each halo only has a single central black hole, so this curve has an imposed upper bound of $\langle N_{\rm{BH,cen}} \rangle \: \le 1$).  At $z = 0$ in particular, we note that $\langle N_{\rm{BH,cen}} \rangle$ actually decreases in the highest-mass halos even as $\langle N_{\rm{BH}} \rangle$ increases, telling us that in the most massive halos, the central black hole is often not the most luminous; instead the central black hole has been quenched, while satellite black holes continue to grow more efficiently. We have also used vertical lines to mark the host mass above which 90\%, 75\%, and 50\% of black holes are found.  We note that, due to the slope of the halo mass function, most black holes are found in halos small enough to have $\langle N_{\rm{BH}} \rangle < 1$ but which are much more common than the more massive halos.

\begin{figure}
    \centering
    \includegraphics[width=8.5cm]{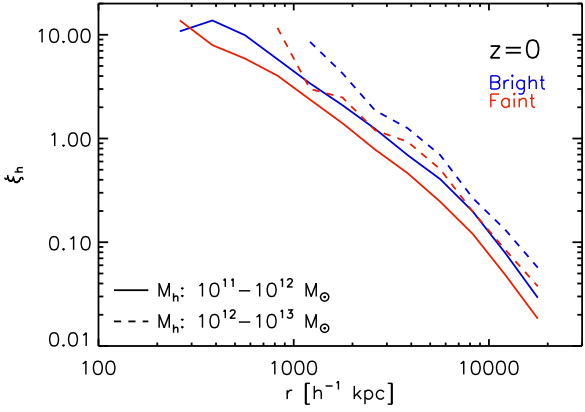}
    \caption{Halo correlation function for halos hosting bright (blue lines) and faint (red lines) AGN in halo mass ranges of $10^{11}-10^{12} M_\odot$ (solid lines) and $10^{12}-10^{13} M_\odot$ (dashed lines).  For each halo mass bin we select the top and bottom quartile in AGN luminosity to form our bright and faint subsamples.  This shows that halos hosting bright AGN tend to be more strongly clustered than those with only faint AGN.}
    \label{fig:haloclustering_bhlum}
\end{figure}

In the middle panel of Figure \ref{fig:hostevolution} we compensate for this by only considering the central black hole in any given halo, which improves the agreement though some discrepancy remains.  To show this more explicitly, in the bottom panel we show the ratio of bias-predicted mass to the actual host mass for the full black hole population (solid lines) and for central black holes only (dashed lines).  At both high redshifts and high luminosities the full and central populations agree with one another, since high redshift and high luminosity black holes tend to be centrals.  At low redshift and moderate- to low- luminosities, satellite black holes play a larger role and so considering only central black holes decreases the discrepancy by up to a factor of 2, but does not remove it entirely (discrepancies up to $M_{\rm{host, bias}}/M_{\rm{host,actual}} \sim 2$). This suggests that halos hosting massive, luminous black holes tend to be more strongly clustered than an equivalent halo mass selected sample.  We test this explicitly in Figure \ref{fig:haloclustering_bhlum}, where we show the halo correlation length in two mass bins, separated into those hosting luminous (blue lines) and faint (red lines) black holes.  For each host mass bin, we select the 25\% of halos hosting the most luminous black holes for our bright sample, and the 25\% with the least luminous black holes for our faint sample.  Here we clearly see that for a given mass range halos hosting brighter AGN tend to cluster more strongly than those with faint AGN.  In fact, we find that the $10^{11}-10^{12} M_\odot$ halos with the brightest AGN are as strongly clustered as the $10^{12}-10^{13} M_\odot$ with the faintest AGN, despite being significantly smaller (note that the extension of $\xi_h$ to smaller scales characterizes the radial extent of the halos, in addition to the mass ranges selected).  This suggests that observational estimates for typical host halo masses may overestimate by a factor of $\sim$2, especially at intermediate redshifts where quasars tend to be most active.

We also consider host growth rates in Figure \ref{fig:hostevolution} by adding an expected halo growth curve (dotted lines) according to the median halo growth rate of \citet{Fakhouri2010}, matched to the redshift at which the growth of $\langle \rm{log} (M_{\rm{host}}) \rangle$ begins growing faster than the expected median growth rate.  We note that our typical host mass does not evolve as a typically growing halo; instead the growth is significantly slower at high-redshift, and faster at low-redshift.  At high redshift, this is largely due to recently seeded black holes: although individual halos hosting black holes are growing, continued seeding of black holes means that new, low-mass halos are being added, partially compensating for the growth of the older black holes.  At lower redshift, however, a larger black hole population combined with rarer black hole seeding minimizes this effect.

At low redshift, we find that the typical host halo increases faster than the expected median halo growth rate.  Here, typical black hole luminosity decreases with time \citep{Sijacki2015}; thus for a given luminosity threshold, as time passes only the most extreme objects continue to satisfy $L_{\rm{BH}} > L_{\rm{BH,min}}$.  In other words, at low redshifts typical $L_{\rm{BH}}$ decreases, and so the smaller halos no longer satisfy the luminosity criterion, leading to a faster rise in $\langle M_{\rm{host}} \rangle$ than the typical halo actually grows.

\begin{figure*}
    \centering
    \includegraphics[width=16.0cm]{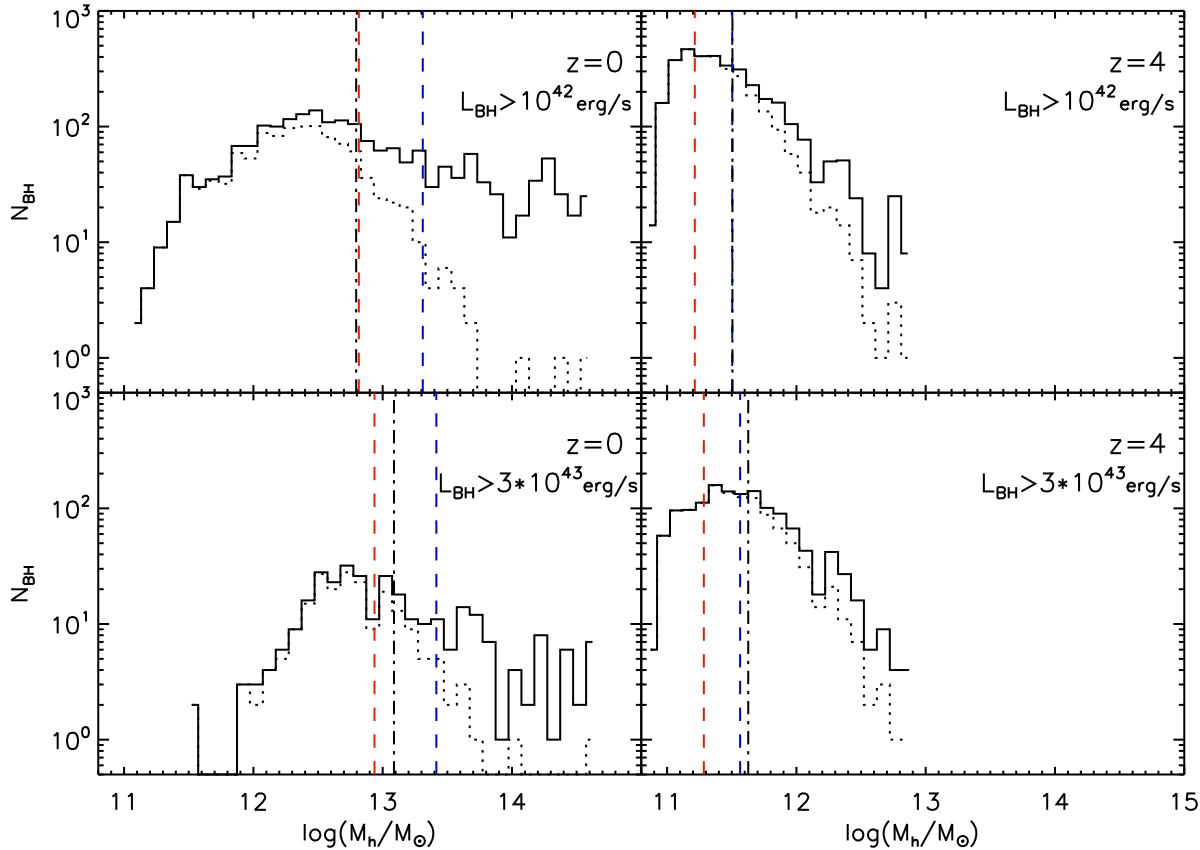}
    \caption{Distribution of host halo masses for black holes above $10^{42}$ erg s$^{-1}$ (top) and $3 \times 10^{43}$ erg s$^{-1}$ (bottom) at z = 0 (left) and z = 4 (right) for full black hole population (solid histogram) and central black hole population (dotted histogram).  Dot-dashed black line shows mean host mass.  Dashed blue shows the predicted mass based on halo clustering.  Dashed red shows the predicted minimum mass based on the bias parameter. The distribution of host halo masses extends below the predicted minimum mass (see Equation \ref{eq:minmass}), with a more significant low-end tail at low-redshift.}
    \label{fig:hostdistribution}
\end{figure*}

To more fully characterize typical host halos, in Figure \ref{fig:hostdistribution} we plot the distribution of halo masses hosting black holes above $L_{\rm{BH}} > 10^{42}$ and $3 \times 10^{43}$ at $z$ = 0 and 4.  The solid histograms show the number of black holes above the given luminosity at each halo mass, while the dotted histogram shows the number of central black holes.  Vertical lines show the mean halo mass (dotted black line) and the typical mass predicted by the black hole bias parameter (dashed blue line).  Consistent with Figure \ref{fig:hostevolution} we see that at high redshift the bias-predicted mass closely matches the actual mean mass.  At low redshift, we note that the distribution of host masses increases significantly, and satellite black holes have a strong impact on the high-mass end of the distribution.  

One of the calculations sometimes made when interpreting observational data is to estimate the minimum mass of a quasar-hosting halo by considering the mass-averaged halo bias:
\begin{equation}
\begin{split}
b_{\rm{BH}}( > L_{\rm{BH,min}}) &= b_h(M_h > M_{h,\rm{min}}) \\
&= \frac{\int_{M_{\rm{h,min}}}^\infty b(M) \frac{dn}{dM} dM}{\int_{M_{h,\rm{min}}}^\infty \frac{dn}{dM} dM} \: .
\end{split}
\label{eq:minmass}
\end{equation}
However, this only holds if a black hole above $L_{\rm{BH,min}}$ is equally likely to be found in any halo above $M_{\rm{h,min}}$, which Figure \ref{fig:occupationnumber} has shown to be inaccurate.  To further characterize the validity of this, we overplot $M_{\rm{h,min}}$ estimated from Equation \ref{eq:minmass} as a dashed red line in Figure \ref{fig:hostdistribution}.  At high redshift this method slightly overpredicts the minimum halo mass, but is relatively close.  At low redshift, however, this substantially overestimates the number of luminous black holes in low-mass halos: although the fraction of low-mass halos hosting luminous black holes is low, the slope of the halo mass function is such that a large fraction of black holes above a given $L_{\rm{BH,min}}$ are nonetheless found in relatively low-mass halos.

\begin{figure}
    \centering
    \includegraphics[width=8.5cm]{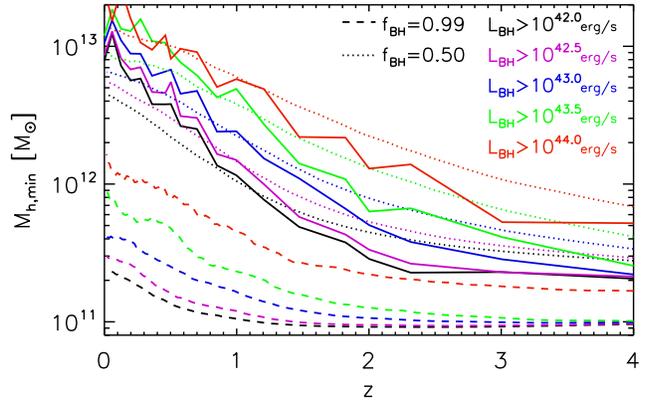}
    \caption{Minimum host mass for several luminosity thresholds calculated using black hole bias (solid lines) and the minimum halo mass above which 99\% (dashed lines) and 50\% (dotted lines) of black holes are found.}
    \label{fig:minmass}
\end{figure}

\begin{figure*}
    \centering
    \includegraphics[width=15cm]{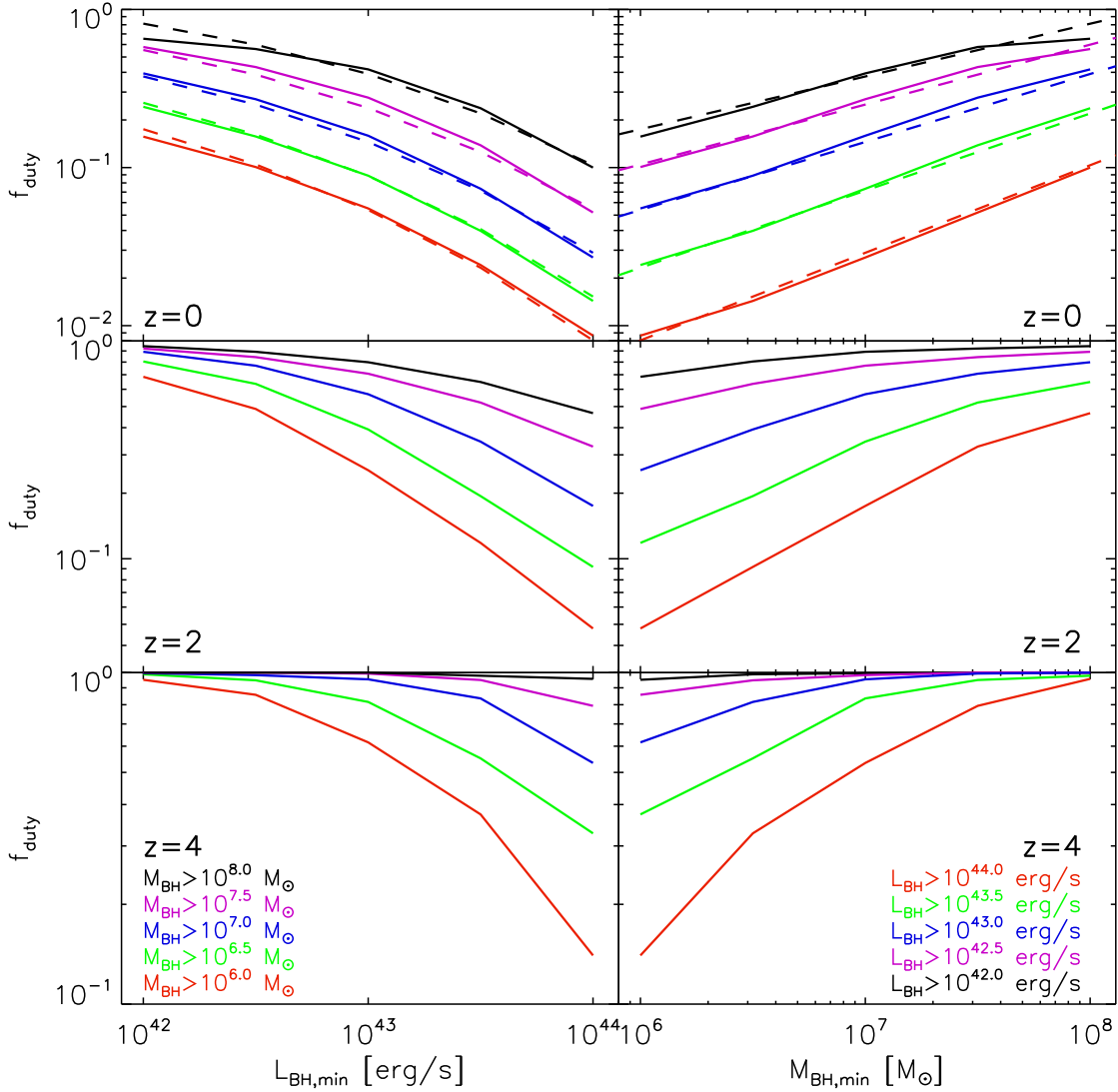}
    \caption{Black hole duty cycle for black holes above a specified minimum mass as a function of $L_{\rm{BH}}$ (left panels) and specified minimum luminosity as a function of $M_{\rm{BH}}$ (right panels), for $z$ = 0, 2 and 4 (top, middle, bottom). The $z = 0$ panels also include the best fitting relation from Equations \ref{eq:powerlaw} and \ref{eq:parameterfit}.  Higher redshift fits are not provided, as the duty cycle rapidly approaches the upper limit of $f_{\rm{duty}} = 1$, and thus diverges from a well-fit power law fit (see text for more details). Note that the y-axis range evolves with redshift, matching the range spanned by our simulation.}
    \label{fig:bhdutycycle}
\end{figure*}

\begin{figure}
    \centering
    \includegraphics[width=8.0cm]{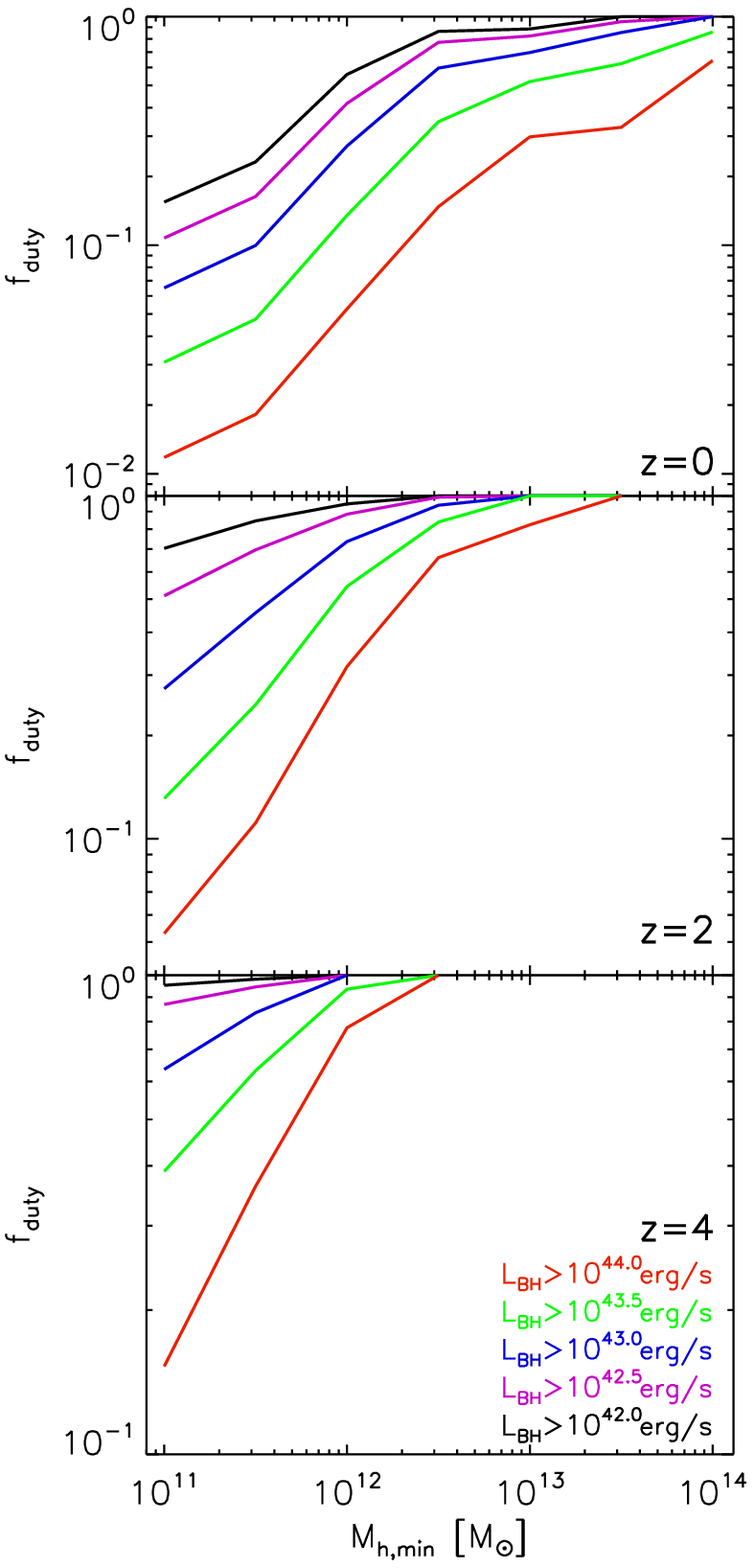}
    \caption{Duty cycle for halos above a given luminosity (line colour) as a function of minimum halo mass, for $z$ = 0, 2 and 4 (top, middle, bottom).  Note that the y-axis range evolves with redshift, matching the range spanned by our simulation.}
    \label{fig:halodutycycle}
\end{figure}

Figure \ref{fig:minmass} shows the evolution of the minimum mass calculated based on the bias parameter as in Equation \ref{eq:minmass} (solid lines).  We also plot the halo mass above which halos are found to host 99\% of black holes (dashed lines) and 50\% of black holes (dotted lines).  Similar to the curves in Figure \ref{fig:hostevolution}, we find that the minimum host mass predicted based on the clustering bias significantly overestimates the actual minimum mass, and is in fact closer to the median mass of host halos, rather than the minimum.  As discussed earlier, this is due to the AGN clustering more strongly than typical halos of equivalent masses.

\subsection{Duty cycle}
\label{sec:dutycycle}

In addition to the clustering properties, we also consider the duty cycle of black holes in the simulation, and the problems with using clustering behaviour to estimate it.  Rather than calculate the fraction of time a given black hole spends above a given luminosity cut, we instead use 
\begin{equation}
f_{\rm{duty}}= \frac{N_{\rm{obj}} (L_{\rm{BH,min}} > L_{\rm{cut}})}{N_{\rm{obj}}},
\label{eq:fduty}
\end{equation}
which provides us with two main advantages: first, it is based on a single snapshot and thus tracking black holes through mergers does not complicate behaviour, and second we can use the same approach to determine both black hole and halo duty cycles (i.e. $N_{\rm{obj}}=N_{\rm{BH}}$ and $N_{\rm{halo}}$).

The black hole duty cycle is plotted in Figure \ref{fig:bhdutycycle} at $z$ = 0, 2 and 4, showing the dependence on both black hole mass and luminosity.  In addition to increasing for lower $L_{\rm{BH,min}}$ (a necessary result of Equation \ref{eq:fduty}), more massive black holes tend to have higher duty cycles, as more massive black holes typically have higher accretion rates.  At $z = 0$ we find the duty cycle to behave very regularly, with difference $L_{\rm{BH,min}}$ and $M_{\rm{BH,min}}$ thresholds tending to only change the normalization of the duty cycle curve, which are well fit by a power law in $M_{\rm{BH,min}}$.

We have fit the black hole duty cycle with a power law form of 
\begin{equation}
f_{\rm{duty}} = 0.1 \times \left ( \frac{M_{\rm{BH}}}{M_0 \left (L_{\rm{BH,min}} \right )} \right )^{\alpha \left (L_{\rm{BH,min}} \right )},
\label{eq:powerlaw}
\end{equation}
where $M_0$ gives us the typical black hole mass at which the duty cycle is 10\%, and $\alpha$ characterizes the sensitivity of $f_{\rm{duty}}$ on $M_{\rm{BH}}$.  Both $M_0$ and $\alpha$ are dependent on the cut used for $L_{\rm{BH}}$, and are also well-fit by power laws:
\begin{equation}
X = A_{X} \times \left ( \frac{L_{\rm{BH,min}}}{10^{43} \rm{erg \: s^{-1}}} \right )^{\beta_{X}},
\label{eq:parameterfit}
\end{equation}
with $A_{M_0} = 4.2 \times 10^6$, $\beta_{M_0}=1.35$, $A_\alpha=0.433$, and $\beta_{\alpha}=0.111$ at $z$ = 0.  We have overplotted these fits as dashed lines in the top panels of Figure \ref{fig:bhdutycycle}, showing excellent agreement.  We emphasize that the dashed lines are a single fit over both $M_{\rm{BH}}$ and $L_{\rm{BH}}$ using Equations \ref{eq:powerlaw} and \ref{eq:parameterfit}, rather than individual fits for each curve.  The only discrepancy occurs when the duty cycle increases to above $\sim$50\%, where $f_{\rm{duty}}$ diverges from a power law as it approaches the maximum possible value of $f_{\rm{duty}}=1$.  The divergence from a power law is more apparent at higher redshifts: $f_{\rm{duty}}$ vs. $M_{\rm{BH}}$ is still a rough power law for $f_{\rm{duty}}$ below $\sim$0.5, but the higher accretion rates at high $z$ \citep[see][]{Sijacki2015} produce high duty cycles across all scales.  For this reason we caution that our fitting function for the duty cycle should only be used below $f_{\rm{duty}} \sim 0.5$ (and at $z = 0$), but this covers the regime of interest when studying duty cycles.  

In Figure \ref{fig:halodutycycle} we plot the halo duty cycle, rather than the black hole duty cycle, i.e. $N_{\rm{obj}}=N_{\rm{halo}}$ in Equation \ref{eq:fduty}.  In this case, $N_{\rm{halo}} (L_{\rm{BH,min}})$ is the number of halos which host at least one black hole above $L_{\rm{BH,min}}$, so $f_{\rm{duty}}$ provides us with a fraction of halos which are present above a specified AGN luminosity.  As expected, duty cycle increases with halo mass, though we note that the relation tends to flatten out with $M_{\rm{halo}}$ more rapidly than with $M_{\rm{BH}}$.

\begin{figure*}
    \centering
     \includegraphics[width=17cm]{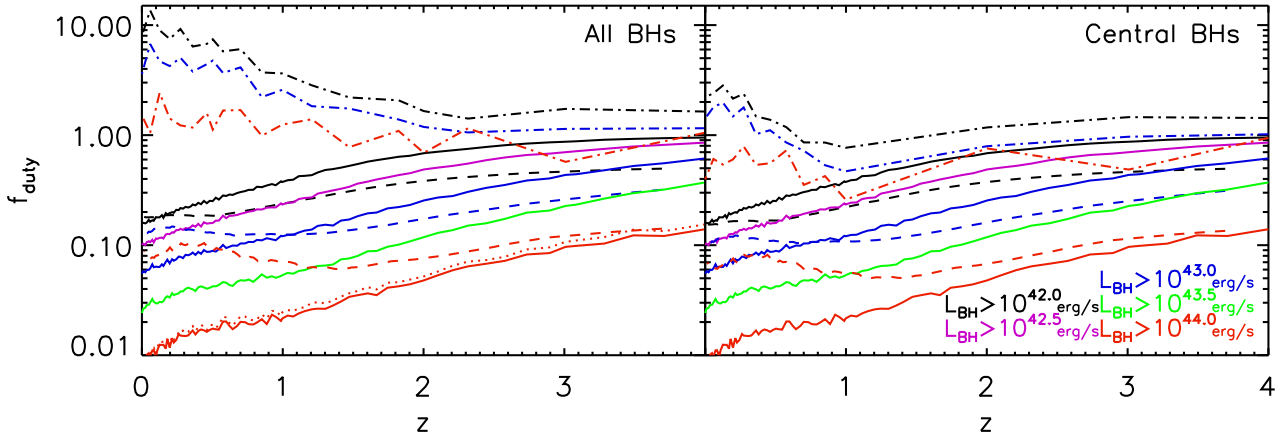}
    \caption{Duty cycle as a function of redshift for varying minimum $L_{\rm{BH}}$ (line colours, as in Figures \ref{fig:bhdutycycle} \& \ref{fig:halodutycycle}). \textit{Solid lines:} Black hole duty cycle for $M_{\rm{BH}} > 10^6 M_\odot$.  \textit{Dashed lines:} Halo duty cycle for halos above a minimum host mass such that 99\% of back holes are included. \textit{Dot-dashed lines:} Bias-predicted duty cycle for all black holes (left), and for central black holes only (right). We find the black hole duty cycle is virtually identical to the halo duty cycle for $M_h > 10^{11.2} M_\odot$ (see pink dotted line for example), but quite poorly matches the halo duty cycle above a minimum mass determined by either the black hole bias or directly from the distribution of host halo masses.} 
    \label{fig:dutycycle_evolution}
\end{figure*}

We compare different methods of calculating duty cycle across cosmic time in Figure \ref{fig:dutycycle_evolution}.  Solid lines show the black hole duty cycle (as in Figure \ref{fig:bhdutycycle}) for a range of lower limits on $L_{\rm{BH}}$, showing the expected decrease in $f_{\rm{duty}}$ with time.  Dotted lines show the halo duty cycle (as in Figure \ref{fig:halodutycycle}) for halos above $10^{11.2}M_\odot$, representing a characteristic minimum mass for black hole occupation in our simulation.  The $10^{11.2} M_\odot$ threshold was selected to provide a close fit to the black hole duty cycle; increasing the halo mass threshold increases the duty cycle, as seen in Figure \ref{fig:halodutycycle}.  We also consider the duty cycle based on the clustering properties, using the equation 
\begin{equation}
\centering
f_{\rm{duty}} = \frac{\int_{L_{\rm{BH,min}}}^\infty \Phi (L) dL}{\int_{M_{h,\rm{min}}}^\infty \frac{dn}{dM} dM} = \frac{N_{\rm{BH}} (L_{\rm{BH}} > L_{\rm{BH,min}})}{N_h (M_h > M_{h,\rm{min}})}  \: ,
\label{eq:fduty_frombias}
\end{equation}
\citep[see, e.g.,][]{MartiniWeinberg2001, Eftekharzadeh2015}, which implicitly assumes that all AGN above $L_{\rm{BH,min}}$ are found in halos above $M_{\rm{h,min}}$, but no other dependency on halo mass.  $L_{\rm{BH,min}}$ is the threshold luminosity used for black holes selction; $M_{\rm{h,min}}$ is the minimum halo mass considered, $\Phi (L)$ is the AGN luminosity function, and $\frac{dn}{dM}$ is the halo mass function.  Of particular importance is $M_{\rm{h,min}}$, which is calculated based on the clustering bias as in Equation \ref{eq:minmass}.  An overestimate of the clustering amplitude would thus produce a correspondingly overestimated $M_{\rm{h,min}}$, and therefore also overestimate the duty cycle.   As shown in Figure \ref{fig:occupationnumber}, the mean occupation number evolves significantly with halo mass contrary to this assumption, suggesting a significant bias between the predicted duty cycle from Equation \ref{eq:fduty_frombias} and the `true' duty cycle.  We plot the estimate from Equation \ref{eq:fduty_frombias} using the full black hole sample as dot-dashed lines in the left panel of Figure \ref{fig:dutycycle_evolution}.  We note two main issues here: the duty cycle is generally larger than 1, and the redshift evolution is significantly different from the `true' black hole duty cycle.  

The main factor contributing to $f_{duty} > 1$ is that individual halos can host multiple AGN above $L_{\rm{BH,min}}$, as shown in Figure \ref{fig:occupationnumber}, while Equation \ref{eq:fduty_frombias} assumes a maximum of one per halo.  We account for this in the right panel of Figure \ref{fig:dutycycle_evolution} by only including central black holes, which decreases $f_{duty}$ but still has $f_{duty} > 1$, due to the mis-estimate of $M_{\rm{h,min}}$.  As shown in Figure \ref{fig:minmass}, calculating $M_{\rm{h,min}}$ from the black hole bias overpredicts the minimum halo mass.  This overestimate means numerous halos are neglected in the denominator of Equation \ref{eq:fduty_frombias}, which overestimates the duty cycle.  By re-calculating $M_{\rm{h,min}}$ such that 99\% of black holes are included (as discussed in Section \ref{sec:hosthalo} and Figure \ref{fig:minmass}), we get the dashed line in Figure \ref{fig:dutycycle_evolution}.  This corrected calculation shows a more reasonable duty cycle, but which still evolves very differently from the actual black hole duty cycle (solid lines).

\begin{figure}
    \centering
    \includegraphics[width=8.5cm]{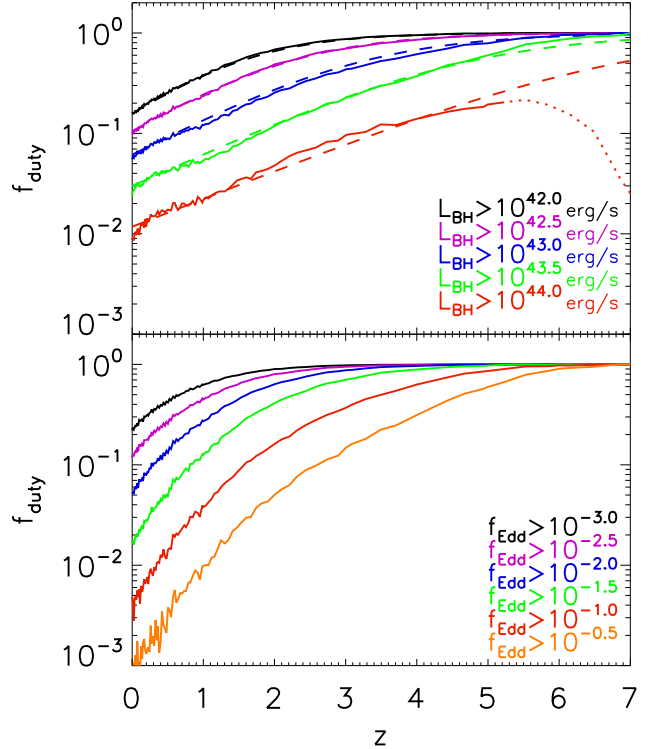}
    \caption{\textit{Top:} Redshift evolution of black hole duty cycle (solid lines; as in Figure \ref{fig:dutycycle_evolution}), together with our redshift-evolution fit from Equation \ref{eq:logistic} (dashed lines).  Note that the fit is performed for $0 \le z \le 4$, but we plot a larger $z$-range to show behaviour at higher redshifts.  The high-redshift decline for the $L_{\rm{BH}} > 10^{44}$ erg s$^{-1}$ population is due to the lack of sufficiently massive black holes capable of radiating at such high rates (with the imposed Eddington limit), represented by a dotted line.  \textit{Bottom:} Redshift evolution for duty cycles of Eddington fraction-selected black hole populations, which eliminates the high-redshift decline.} 
    \label{fig:dutycycle_evolution_fit}
\end{figure}

We also fit the redshift evolution of the duty cycle of $M_{\rm{BH}} > 10^6 M_\odot$ black holes to a logistic function
\begin{equation}
f_{\rm{duty}}=\frac{1}{1+e^{-k(z-z_0)}},
\label{eq:logistic}
\end{equation}
where $k$ and $z_0$ are both found to be well-fit by power law-functions in $L_{\rm{BH,min}}$, as in Equation \ref{eq:parameterfit}, with $A_k = 0.87$, $\beta_k=-0.127 $, $A_{z_0}= 3.13$, and $\beta_{z_0} = 0.338$.  The top panel of Figure \ref{fig:dutycycle_evolution_fit} shows this fit as dashed lines, demonstrating that the fits are excellent across a full range of redshifts and luminosities.  Although the fitting was performed over the range $0 < z < 4$, we have extended the plotting range of Figure \ref{fig:dutycycle_evolution_fit} to higher redshift.  For $L_{\rm{BH,min}} < 10^{44}$ erg s$^{-1}$, the evolution out to higher redshift remains excellent.  Above $z \sim 5$, there are very few black holes which have grown large enough to reach $10^{44}$ erg s$^{-1}$, due to the imposed Eddington limit of the simulation (see Section \ref{sec:numerical}).  Thus we expect the duty cycle for the highest luminosity to drop at high-redshift, simply due to the lack of sufficiently massive black holes, and the fitting function should not be used.  We plot the $L_{\rm{BH}} > 10^{44}$ erg s$^{-1}$ curve as a dotted line for redshifts at which the fraction of black holes capable of reaching $10^{44}$ erg s$^{-1}$ at maximum (i.e. Eddington) accretion is less than the predicted duty cycle (dashed line).  To confirm this explanation, the bottom panel shows the duty cycle for black hole populations selected by Eddington fraction rather than luminosity, finding the expected smooth increase with redshift.  Here we show that the majority of black holes do approach the Eddington limit for $z > 5$, and so the decrease in the $L > 10^{44}$ erg s$^{-1}$ curve in the upper panel is indeed due to limitations on the mass function at high-redshift rather than a change in active fraction.

\subsection{Luminosity-dependent cross-correlation with satellite galaxies}
\label{sec:crosscorrelation}

\begin{figure}
    \centering
    \includegraphics[width=8.0cm]{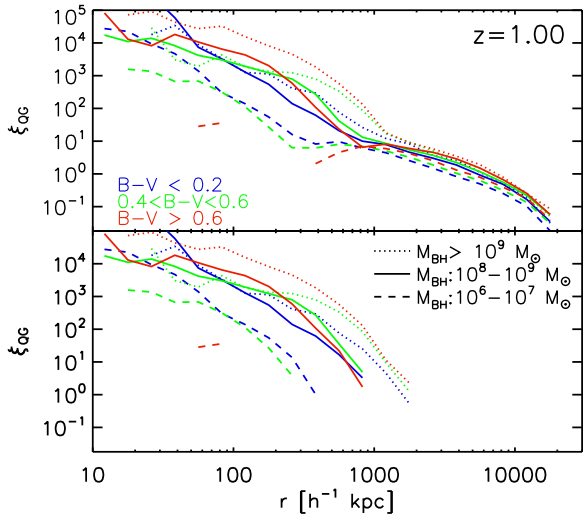}
    \caption{\textit{Top:} Cross correlation between black holes selected by mass (to characterize integrated black hole feedback) and galaxies selected by B-V colour (to characterize galaxy quenching).  \textit{Bottom}: Cross-correlation function 1-halo term only. The top panel shows the cross-correlation of black holes and quenched galaxies is strongly $M_{\rm{BH}}$-dependent.  The lower panel, however, shows that this dependence is actually due to the radial extent of the host halo, which also correlates with $M_{\rm{BH}}$ (also see Figure \ref{fig:crosscorrelation_rvir}). }
    \label{fig:crosscorrelation}
\end{figure}

One final aspect of our analysis is to use AGN clustering to look for signals of AGN-induced galaxy quenching.  Rather than use the black hole autocorrelation used in the rest of our work, here we use the cross-correlation function between black holes and galaxies, with a particular emphasis on satellite galaxies.  In particular, if AGN were capable of quenching star formation in satellite galaxies, we would expect to find quenched galaxies to be preferentially found near massive (or strongly-accreting) black holes, which should be detectable in the black hole - galaxy cross-correlation function.  To investigate this, we used many different selection criteria for both black holes and galaxies, including black hole mass and luminosity, galaxy mass, stellar luminosity, galaxy colour, star formation rate, etc.  The most promising calculation used black holes selected by mass (representing an integrated accretion history) and galaxies selected by colour (characterizing the degree to which star formation has been quenched, while remaining less sensitive to short timescale variations than specific star formation rate).  The top panel of Figure \ref{fig:crosscorrelation} shows the cross correlation for these selections, illustrating that the 1-halo clustering signal (below $\sim 1000 \: h^{-1}$ kpc) is strongest for massive ($> 10^{9} M_\odot$) black holes and quenched galaxies ($B-V > 0.6$).  In particular, we note that the $B-V < 0.2$ and $0.4 < B-V < 0.6$ behave similar to one another regardless of central black hole mass, while the $B-V > 0.6$ galaxies tend to be more strongly clustered about the most massive black holes,  suggesting a possible connection between total energy radiated by the black hole and quenched satellites.  

\begin{figure}
    \centering
    \includegraphics[width=8.0cm]{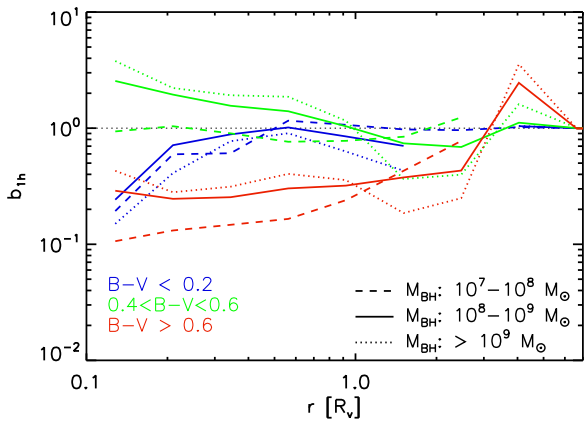}
    \caption{1-halo bias ($b_{\rm{1h}} = \sqrt{\xi_{QG}/\xi_{GG}}$) between black hole-galaxy cross-correlation ($\xi_{QG}$) and galaxy autocorrelation ($\xi_{GG}$), as a function of host virial radius.  Demonstrates no $M_{\rm{BH}}$-dependence in cross-correlation with quenched galaxies, when controlling for host halo size.}
    \label{fig:crosscorrelation_rvir}
\end{figure}

However, further investigation shows that this is not a causal connection between massive black holes and quenched satellite galaxies, but rather a signature of halo radius.  In the bottom panel of Figure \ref{fig:crosscorrelation} we show the 1-halo term only, demonstrating that the difference appears to be largely due to a rescaling of radial separation, with massive black holes tending to be found in galaxies with the largest radial extent.  We take this one step further in Figure \ref{fig:crosscorrelation_rvir}, showing a bias between the cross correlation function ($\xi_{\rm{QG}}$) and the galaxy-galaxy autocorrelation function ($\xi_{\rm{GG}}$; bias defined as $\sqrt{\xi_{\rm{QG}}/\xi_{\rm{GG}}}$), with separation defined in units of the virial radius rather than physical size.  Here we see that, after rescaling based on the host virial radius, there is no significant difference between any samples.  We also considered a possible dependence on the mass of black hole relative to its host halo, and also checked smaller-volume simulations with both stronger and weaker radio-mode feedback (as any causal link should be more apparent when feedback is stronger), and confirmed the lack of any quenching signature.  This suggests that massive central black holes, although capable of quenching their host galaxies \citep[see, e.g.][]{Sijacki2015}, do not tend to quench star formation of satellite galaxies in the same host halo.

\section{Conclusions}
\label{sec:conclusions}

In this work we have used the Illustris simulation to investigate the clustering of supermassive black holes across a range of redshifts and luminosities.  In addition to general agreement with observations, we use the clustering information to link black hole properties to the host masses as well as the AGN duty cycle of black holes and galaxies in general.  Our main conclusions are the following:

\begin{itemize}

\item AGN clustering is found to be luminosity dependent, but primarily at small (1-halo) scales.  At larger scales, luminosity dependence primarily occurs at intermediate redshift, where black hole accretion tends to be strongest.  

\item Correlation length ($r_0$) can have significant luminosity dependence, especially at intermediate redshifts and when satellite black holes are included.  $r_0$ reaches a minimum at $z \sim 1.5-2$, with higher-redshift evolution being strongest for low luminosity thresholds.  Our $r_0$ estimates are generally lower than observational measures.  This is largely due to the limited luminosity range in our simulation (imposed by the simulation volume), however, and adjusting observations to match our mean luminosities produces fully consistent results.

\item Our estimated black hole bias matches observations very well at low-redshift.  For $z > 2$ we predict a lower bias than \citet{Croom2005} and \citet{Shen2009}, but consistent with \citet{Eftekharzadeh2015}.  

\item AGN clustering tends to be stronger than the expected clustering of halos of comparable mass; as a result, AGN hosts tend to be less-massive than predictions made based on AGN clustering.  Although strongest when satellite halos (found most commonly in the largest halos) are included, this effect remains even when only central black holes are considered.  This suggests that typical host halo masses found based on clustering behaviour may be underestimated by a factor of $\sim 2$, especially at intermediate redshifts.  

\item The scatter in black hole-host scaling relations and typical black hole Eddington fractions results in a wide distribution of host halo masses.  Although the distribution for any given $L_{\rm{BH,min}}$ does drop off at low halo mass, there does remain a low-mass tail, especially at low-redshifts.

\item Due to AGN being more strongly clustered than halos matched to the typical hosts and both the wide range and low-end tail of the host halo distribution, estimating the minimum host halo mass from AGN clustering tends to substantially overestimate $M_{\rm{h,min}}$, which can have a strong impact on duty cycle estimates.  

\item At low redshift, the black hole duty cycle follows a power law in $M_{\rm{BH}}$, with a normalization set by the luminosity threshold.  Higher redshifts also tend to follow a rough power law for $f_{\rm{duty}} < 0.5$, above which the curve flattens out.

\item Black hole duty cycle decreases with time, well fit by a logistic function with lower $L_{\rm{BH,min}}$ thresholds decreasing more rapidly and at lower redshifts.  

\item Black hole duty cycle is well matched by the halo duty cycle for halos with $M_h > 10^{11.2} M_\odot$, representing a characteristic minimum mass for black hole occupation.  

\item Estimating the duty cycle from AGN number and expected halo number above a given $M_{\rm{h,min}}$ is very inaccurate.  In addition to the mis-estimate of $M_{\rm{h,min}}$, the rapid growth of typical host halo masses at low-redshift produces a significant increase in the calculation of $f_{\rm{duty}}$ which is not found in the true black hole duty cycle.

\item We used the AGN-galaxy cross-correlation function to look for a possible signature of AGN-induced quenching of satellite galaxies.  Although $\xi_{\rm{QG}}$ does show $M_{\rm{BH}}$-dependent clustering of quenched galaxies, we find this signal is caused by the larger physical size of halos hosting massive black holes rather than a direct causal link.  After controlling for halo size, we find no evidence for AGN inducing quenching in satellite galaxies.

\end{itemize}

Using the Illustris simulation, we have shown black hole and AGN clustering consistent with current observations, and characterized the luminosity dependence of AGN clustering, which is strongest at intermediate redshift ($z \sim 1.5-2$).  One of the most important aspects of clustering analysis is the use of a clustering signal to characterize properties of the host halos, particularly the halo mass.  We find that the typical approach taken (matching AGN clustering to analytic estimates for halo clustering) does very well at high-redshift, but can overestimate host mass by $\sim 50 \%$ at low-redshift, as low-redshift AGN are found to cluster more strongly than an equivalent-mass halo.  Finally, we considered the use of AGN clustering as an estimator for black hole duty cycles.  A typical method for this estimation is to assume a minimum host mass for a given AGN luminosity (see Equation \ref{eq:minmass}) and a constant duty cycle among halos above this threshold.  Contrary to this assumption, however, we find a wide distribution of halo masses, including a low-mass tail. This scatter among host masses (as also found in other simulations) must be accounted for or the AGN duty cycle can be strongly overestimated, particularly at low-redshift.  Overall, we find the black hole duty cycle to evolve smoothly with redshift, and we provide numerical fits characterizing this evolution as well as the dependence on black hole mass and AGN luminosity.

In summary, our work highlights that while black hole clustering is a powerful probe of host halo properties, cosmological simulations, such as Illustris, are needed to fully characterize and account for a number of biases which would otherwise lead to systematically overestimated clustering-predicted host halo masses and black hole duty cycles.

\section*{Acknowledgments}

The authors would like to thank Martin Haehnelt, Volker Springel, and Lars Hernquist for their useful comments on this work, and the referee for a very constructive report.  CD and DS acknowledge support by the ERC starting grant 638707 ``Black holes and their host galaxies: co-evolution across cosmic time.'' DS further acknowledges support from the STFC.  Simulations were run on the Harvard Odyssey and CfA/ITC clusters, the Ranger and Stampede supercomputers at the Texas Advanced Computing Center as part of XSEDE, the Kraken supercomputer at Oak Ridge National Laboratory as part of XSEDE, the CURIE supercomputer at CEA/France as part of PRACE project RA0844, and the SuperMUC computer at the Leibniz Computing Center, as part of project pr85je.

 \bibliographystyle{mn2e}       
 \bibliography{astrobibl}       

\end{document}